\documentclass[prd,aps,showpacs,12pt,amssymb,preprintnumbers,amsmath,amsfonts,nofootinbib]{revtex4-1}

\usepackage{latexsym}
\usepackage[latin1]{inputenc}
\usepackage{amsmath}
\usepackage{mathtools}
\usepackage[dvips]{graphicx}
\usepackage{hyperref}  
\usepackage{dcolumn}                                   
\usepackage{color}
\usepackage{slashed}
\usepackage{ulem}
\usepackage{braket}	

\newcommand{\eps}{{\epsilon}}

\newcommand{\pa}{\partial}
\newcommand{\g}{\gamma}
\newcommand{\q}{\boldsymbol{q}}

\newcommand{\bv}{\boldsymbol{v}}

\begin{document}
\date{\today}
\title{Photon quantum kinetic equations \\
and collective modes in an axion background} 

\author{Marc Comadran${}^{1,2,3}$}
\email{mcomadca60@alumnes.ub.edu}

\author{Cristina Manuel${}^{1,2}$}
\email{cmanuel@ice.csic.es}

\affiliation{${}^1$Instituto de Ciencias del Espacio (ICE, CSIC) \\
C. Can Magrans s.n., 08193 Cerdanyola del Vall\`es,  Spain 
and 
\\
${}^2$Institut d'Estudis Espacials de Catalunya (IEEC), Ed. RDIT, Campus UPC, 08860 Castelldefels (Barcelona), Spain 
and
\\
${}^3$Departament de F\'isica Qu\`antica i Astrof\'isica 
and Institut de Ci\`encies del Cosmos, Universitat de Barcelona (IEEC-UB), Mart\'i i Franqu\'es 1, 08028 Barcelona,  Spain}

\begin{abstract}
We develop a quantum kinetic theory for photons in the presence of an axion background and in the collisioness limit. In deriving the classical regime of our quantum kinetic equations, we observe that they capture well known features of axion electrodynamics. By projecting the Wigner function onto a polarization basis, relating the Wigner matrix function with the Stokes parameters, we establish the  dispersion relations and transport equations for each polarization space component. Additionally, we investigate how the axion background affects the dispersion relations of photon collective modes within an electron-positron plasma at equilibrium temperature $T$. While the plasmon remains unaffected, we find that the axion background breaks the degeneracy of transverse collective modes at order $eg_{a\g}T(\pa a)$, where $e$ represents the electron charge, $g_{a\g}$ denotes the photon-axion coupling, and $\pa a$ represents the scale associated with variations in the axion field. 
\end{abstract}

\maketitle

\section{Introduction}

The exploration of axion electrodynamics holds interest in various domains of physics.  Originally
conceived within the realm of high-energy physics to address the so called CP problem of QCD \cite{Peccei:1977hh,Peccei:1977ur,Weinberg:1977ma,Wilczek:1977pj},
the term axion is nowadays  more generically used to refer to a broader class of light
pseudoscalar particles, regardless or not they are related to QCD. Axions naturally manifest in  extensions of the Standard Model of particle physics, thus deserving serious consideration as potential candidates for dark matter.
There are several intense experimental
programs to search for these elusive particles, both in the laboratory an in astrophysical scenarios
(see, for example, \cite{Adams:2022pbo,Caputo:2024oqc}, and references therein). Concurrently, analogous axion-photon couplings manifest in certain condensed matter systems \cite{Sekine:2020ixs}, giving account of topological
magnetoelectric phenomena. 

In this article we will consider photon properties in a plasma when there is also an axion background. A plasma is typically characterized by either a temperature $T$ and/or chemical potential $\mu$. We will consider that the axion wavelength is much lesser than any of these scales $T, \mu$ which describe the medium.  The interactions between axions and photons are described by the Lagrangian
\begin{equation}\label{L_photon+axion}
\mathcal{L}_{\text{int}}=\dfrac{1}{4}g_{a\g}a F_{\mu\nu}\Tilde{F}^{\mu\nu} \ .
\end{equation}
Here $\Tilde{F}^{\mu\nu}=\dfrac{1}{2}\eps^{\mu\nu\alpha\beta}F_{\alpha\beta}$ is the dual of the electromagnetic tensor $F^{\mu\nu}=\pa^\mu A^\nu-\pa^\nu A^\mu$, being $\eps^{\mu\nu\rho\sigma}$ the Levi-Civita tensor, the fields $A^{\mu}(x)$ and $a(x)$ are associated to photons and axions respectively, and $g_{a\g}$ is the axion-photon coupling \cite{Caputo:2024oqc}. This coupling suggests that an axion background acts as an effective chiral medium. There has been a renewed interest in the last decade on chiral media, with the discovery of a variety of new quantum chiral transport phenomena \cite{Kharzeev:2013ffa}, such as the chiral magnetic effect. A clear parallelism among axion electrodynamics and chiral media has been stressed \cite{Qiu:2016hzd}. 

Quantum field theory methods have been developed for the study of relativistic plasmas \cite{Bellac:2011kqa}. It is by now well understood that the particle fields of different  energy scales have to be treated differently (see, for example \cite{Litim:2001db}). Effective field theories have been designed for that purpose. For momenta scales of the order of the
temperature $T$ or higher, the photons are treated as quasiparticles, which are more efficiently described with transport equations. For momenta much lower than the temperature $T$ the photon fields are then described as classical background fields, whose properties  are then modified by the medium. Collective modes then emerge
for those scales, with the appearance of the so called plasmon mode \cite{Silin:1960pya,Weldon:1982aq}, which is absent in vacuum.

The purpose of this article is to study how an axion background modifies both the photon transport equation and also the collective modes of a thermal plasma.  We will assume that the time and spatial scales of variation of the axion are much less than the photon momentum. While
our study would be valid for astrophysical and cosmological plasmas, it can be also of used for other condensed matter systems.
We will use  well established quantum field theory methods to study these two effects. Similar photon transport equations have  been derived  \cite{Shakeri:2022usk,McDonald:2023ohd}. We will comment later on differences with Ref.~\cite{Shakeri:2022usk}, while in Ref.~\cite{McDonald:2023ohd} the axions are considered as quasiparticles and not as a background. Therefore there the axions interact with the photons through the collision term, allowing for the conversion of axions into photons in the background of a magnetic field, as those commonly found in astrophysical plasmas. In this work, we ignore these processes, while they could be easily incorporated by including a proper collision term to our transport equations.

The paper is structured as follows; in Sec.~\ref{Sec.II} we derive quantum kinetic equations for photons in the presence of an axion background using the Keldysh Schiwgner formalism and in Sec.~\ref{Sec.III} we address the effect of the axion background onto the collective modes of photons. Finally in Sec.~\ref{Sec.IV} we discuss our findings and make contact with other approaches found in the literature. We also give some details on the derivation of the kinetic equations in Appendix \ref{A}. Appendix \ref{B} is devoted to provide the transport equations in a linear polarization basis.

Let us establish our notations and conventions. We use the signature $\text{diag}(g^{\mu\nu})=(1,-1,-1,-1)$,
and the normalization $\eps^{0123}=-\eps_{0123}=1$ for the Levi-Civita tensor. Rising or lowering spatial indices produces a minus sign e.g if $A^\mu=(A^0,A^i)$ is a vector in Minkowski space, we have $A^i=-A_i$.  Boldface letters will be used to denote three dimensional vectors $A^i=\boldsymbol{A}$. The four gradient is $\pa_\mu=(\pa_0,\pa_i)$, where $\pa_i=\nabla$. The Minkowski product between $A^\mu$ and $B^\mu$ is defined as $A\cdot B= g_{\mu\nu}A^\mu B^\mu = A^0B^0-\boldsymbol{A}\cdot\boldsymbol{B}$. Natural units $ \hbar= c =k_B =1$ are used throughout.

\section{Quantum kinetic equations for photons in an axion background}
\label{Sec.II}
The Keldysh-Schwinger formulation of quantum field theory has become a well-established tool to study relativistic plasmas, whether they are at or far to thermal equilibrium \cite{Danielewicz:1982kk}. In the Keldysh-Schwigner formalism, non-equilibrium Green functions are defined as $2\times 2$ matrix in the complex, closed time path contour, see e.g Ref.~\cite{Mrowczynski:2016etf} for a recent review. In the case of photons one defines
\begin{equation}\label{propagator_medium}
\boldsymbol{G}^{\mu\nu}(x,y)=
\begin{pmatrix}
G^{c,\mu\nu}(x,y) & G^{<,\mu\nu}(x,y) \\
G^{>,\mu\nu}(x,y) & G^{a,\mu\nu}(x,y)
\end{pmatrix}
=
\begin{pmatrix}
\braket{\mathcal{T} A^\mu(x)A^\nu(y)} & \braket{A^\nu(y)A^\mu(x)} \\
\braket{A^\mu(x)A^\nu(y)} & \braket{\Tilde{\mathcal{T}}A^\mu(x)A^\nu(y)} 
\end{pmatrix}    \ ,
\end{equation}
where $\mathcal{T}$ and $\Tilde{\mathcal{T}}$ denote time and anti-time ordering along the complex path, respectively, and $\braket{\ldots}$ stands for average over an ensemble of states. In order to derive kinetic equations it is sufficient to study the dynamics of the lesser (or  greater) component of the Green function $G^{<,\mu\nu}(x,y)$ (or $G^{>,\mu\nu}(x,y)$), as this is related to the photon phase space density after a Wigner transformation.

A relevant problem that emerges in the present formulation is that 
the components of the Green function in Eq.~\eqref{propagator_medium} are not gauge invariant quantities, and they contain non-transverse degrees of freedom. There are various ways to circumvent this inconvenience, the usual prescription is to impose gauge fixing conditions, although there are other approaches, such as defining a gauge-invariant two-point Green function \cite{Vasak:1987um}. In a forthcoming publication, we will explore alternative possibilities by employing effective field theory techniques.
In this work, we adopt the former and impose the Lorentz gauge $\pa^\mu A_\mu=0$, which leads to the following gauge conditions for the lesser component of the Green function
\begin{subequations}
\begin{gather}\label{Lorenz_gauge_x}
\pa_{x,\mu}G^{<,\mu\nu}(x,y)=0 \ , 
\\    \label{Lorenz_gauge_y}
\pa_{y,\nu}G^{<,\mu\nu}(x,y)=0 \ .
\end{gather}
\end{subequations}
The equations of motion obeyed by each component of the Green function can be deduced from the Kadanoff-Baym equations \cite{Kadanoff:1981}. In the collisionless limit and allowing photons to interact with the axion background through the coupling in Eq.~\eqref{L_photon+axion}, the Kadanoff-Baym equations take the simple form
\begin{subequations}
\begin{align}\label{motion_x}
& \left(g_{\mu\lambda}\square-g_{a\g}\eps_{\mu\lambda\alpha\beta}(\pa^\alpha a)\pa^\beta\right)_x  G^{<,\mu\nu} (x,y)=0 \ ,  
\\ \label{motion_y}
& \left(g_{\mu\nu}\square-g_{a\g}\eps_{\mu\nu\alpha\beta}(\pa^\alpha a)\pa^\beta\right)_y  G^{<,\lambda\mu} (x,y)=0 \ .    
\end{align}
\end{subequations}
Where $\square=\pa^\mu \pa_\mu$ and the suffix $(\ldots)_{x}$ indicates that all operators act on the coordinate $x^\mu$. The second equation above, in which the operators act on $y^\mu$, is easily obtained by considering the equation of motion obeyed by the greater two-point Green function, renaming $x^{\mu}\leftrightarrow y^{\mu}$ and using the property
\begin{equation}\label{property}
G^{>,\nu\mu} (y,x)=G^{<,\mu\nu} (x,y)  \ .  
\end{equation}
The phase space density, also called Wigner function, is defined as the Wigner transform of the Green function
\begin{equation} \label{Wigner_function}
G^{<,\mu\nu} (X,q)=\int d^4s\, e^{iq\cdot s} G^{<,\mu\nu} (X+s/2,X-s/2) \ ,
\end{equation}
where $X^\mu=(x^\mu+y^\mu)/2$ and $s^\mu=x^\mu-y^\mu$ are the center and relative coordinates, respectively.
Let us recall, as said in the introduction,  that we assume that $|\partial_\mu a|/|a| \ll q_\mu$.
Adding and subtracting the Wigner transformed equations of motion of Eqs.~(\ref{motion_x}-\ref{motion_y}) one finds the collisionless equations
\begin{subequations}\label{Kinetic_Equations}
\begin{gather}\label{Dispersion}
\left(q^2-\dfrac{\pa^2}{4}\right)G^{\mu\nu} +\dfrac{g_{a\g}}{2}\left(
\eps^{ \mu \rho  \alpha \beta} A_{\alpha\beta} G_{\rho }^{\ \nu}+\eps^{\nu \rho \alpha \beta} A^*_{\alpha\beta} G^\mu _{ \ \rho }\right)
=0 \ ,   
\\\label{Transport}
(iq\cdot \pa)G^{\mu\nu}  +\dfrac{g_{a\g}}{2} \left(
\eps^{ \mu \rho  \alpha \beta} A_{\alpha\beta} G_{\rho }^{\ \nu} -\eps^{\nu \rho \alpha \beta} A^*_{\alpha\beta} G^\mu _{ \ \rho }\right)
=0 \ ,
\end{gather}
\end{subequations}
which have to be complemented with the Wigner transformed gauge conditions of Eqs.~(\ref{Lorenz_gauge_x}-\ref{Lorenz_gauge_y})
\begin{equation} \label{GF_Wigner_space}
\left( \frac 12 \partial_\mu - i q_\mu \right) G^{\mu \nu} = \left( \frac 12 \partial_\mu + i q_\mu \right) G^{\nu \mu} =0 \ ,
\end{equation}
and the additional condition, necessary to eliminate the residual  ambiguity of the Lorentz gauge
\begin{equation}  \label{gauge_ambiguity}
u_\mu G^{\mu \nu} =u_\mu G^{\nu \mu} =0 \ ,
\end{equation}
where $u^\mu$ is a time-like vector representing the velocity of the reference frame, satisfying $u^2=1$. When writing the above equations, we used the notation $\pa^\mu = \partial/\partial X_\mu $ and dropped the lesser symbol and the arguments of the Wigner function to enlighten the notation, $G^{\mu \nu}=G^{<,\mu \nu}(X,q)$. Additionally, we established
\begin{equation} \label{A_tensor}
A_{\alpha\beta} G_{\rho }^{\ \nu} \equiv  \sum_{n=0}^\infty \frac{ (- i \Delta)^n}{n !}\partial_\alpha a(X) \left(\frac 12 \partial_\beta -i q_\beta \right)G_{\rho }^{\ \nu}(X, q) \ ,
 \end{equation}
being $\Delta = \frac 12 \frac{\partial}{\partial q} \cdot \frac{\partial}{\partial X}$. For details regarding the derivation of the kinetic equations, we refer to Appendix~\hyperref[A]{A}. In the absence of an axion background, we reproduce the photon quantum kinetic equations found in  Refs.~\cite{Hattori:2020gqh,Huang:2020kik,Lin:2021mvw}. The Keldysh-Schwinger formalism could also be used  to reproduce the collision term in the photon transport equation. We will not add a collision term in this reference, but refer to \cite{McDonald:2023ohd} for that purpose.

\subsection{Classical transport equation in the polarization basis}

It is convenient to  obtain the classical limit of the photon transport equations, which are easier to handle and are enough to study the physics at large scales. For that purpose, one 
should carry out a gradient expansion, assuming that variations
of the Wigner function on the scale associated to the center coordinate are much less than those on the relative coordinate i.e $\pa_X^\mu \ll \pa_s^\mu$.
That this corresponds to a classical limit can be seen by going to momentum space in the relative coordinate and restoring $\hbar$, as one then assumes
$ \hbar \pa_X^\mu \ll q^\mu$ when performing the gradient expansion. In a thermal plasma, as most photons have momenta of the order of the temperature $T$, this simply implies to look for variations at scales larger than the inverse of the temperature. Thus, we perform a gradient expansion of Eqs.~(\ref{Dispersion}-\ref{Transport}) and the Wigner function
\begin{equation}
G^{\mu\nu}(X,q)= G^{\mu\nu}_{(0)}(X,q)+ G^{\mu\nu}_{(1)}(X,q)+\mathcal{O}(\pa_X^2) \ ,
\end{equation}
where $G^{\mu\nu}_{(0)}(X,q)$ should be understood as the classical limit of the Wigner function, while higher order terms in the gradient expansion correspond to the quantum corrections. 
As we will only consider  the classical limit in this work, we drop the subindex $(0)$, and understand that we only keep the $G^{\mu\nu}_{(0)}$ in the expansion.
In the classical limit the kinetic equations read
\begin{subequations}\label{classical_KE}
\begin{gather}\label{Dispersion-classical}
q^2G^{\mu\nu} -\dfrac{ig_{a\g}}{2}(\pa_\alpha a)q_\beta\left(
\eps^{ \mu \rho  \alpha \beta}  G_{\rho }^{\ \nu}-\eps^{\nu \rho \alpha \beta}  G^\mu _{ \ \rho }\right)
=0 \ ,   
\\ \label{Transport-classical}
(iq\cdot \pa)G^{\mu\nu}  -\dfrac{ig_{a\g}}{2}(\pa_\alpha a)q_\beta \left(
\eps^{ \mu \rho  \alpha \beta} G_{\rho }^{\ \nu} +\eps^{\nu \rho \alpha \beta} G^\mu _{ \ \rho }\right)
=0 \ ,
\end{gather}
\end{subequations}
and the classical Wigner function is constrained by
\begin{equation}
q_\mu G^{\mu \nu} = q_\mu G^{\nu \mu}=u_\mu G^{\mu \nu} = u_\mu G^{\nu \mu} \ .
\label{classical-Lorentzcondition}
\end{equation}
After imposing the condition that the Wigner function is orthogonal to $u^\mu$, the resulting framework is essentially identical to adopting the Coulomb gauge \cite{Hattori:2020gqh}. Yet another way to eliminate the residual gauge ambiguity of the Lorentz gauge is to project the Wigner function into the physical space, using transverse projectors \cite{Beneke:2010eg}. 

It is convenient to write the transport equation in a polarization basis. If we introduce a two dimensional basis of polarization vectors (e.g $\eps_a^{\mu}$ with $a=\lbrace 1,2\rbrace$) satisfying, $\eps_a^*\cdot \eps_b =\delta_{ab}$, and  $\eps_a \cdot u = \eps_a \cdot q =0 $, then the Wigner function can be expressed as
\begin{equation}\label{Wigner_function_basis}
G^{\mu\nu} (X,q)= \sum_{a,b =1,2} \eps_a^{*\mu} \eps_b^{\nu} \, G^{ab} (X,q) \ . 
\end{equation}
Projecting Eqs.~(\ref{Dispersion-classical}-\ref{Transport-classical}) onto the polarization basis we easily obtain the kinetic equations obeyed by the polarization space components of the Wigner function
\begin{subequations}
\begin{align}
& q^2G^{ab} -\dfrac{ig_{a\g}}{2}\eps_{\mu\nu\alpha\beta}(\pa^{\alpha} a)q^\beta\left(\eps_c^{*\mu} \eps_{a}^{\nu} G^{cb} 
-\eps_c^{\mu}  \eps_{b}^{*\nu} G^{ac} \right)=0 \ ,
\\  
& (iq\cdot \pa)G^{ab}  -\dfrac{ig_{a\g}}{2}\eps_{\mu\nu\alpha\beta} (\pa^{\alpha} a)q^\beta\left(\eps_c^{*\mu} \eps_{a}^{\nu} G^{cb}
+ \eps_c^{\mu}  \eps_{b}^{*\nu} G^{ac} \right)=0 \ ,
\end{align}    
\end{subequations}
where summation over repeated indices should be understood. It is simpler to solve the transport equations in a circular polarization basis, since the polarization space components of the Wigner function then decouple. Explicitly, we introduce the polarization basis vectors $\eps_a^{\mu}=\lbrace \eps_{+}^{\mu},\eps_{-}^\mu\rbrace$, characterized by the properties
\begin{equation}\label{circular}
\eps_{\pm}^*\cdot \eps_{\pm}=0 \ , \quad \eps_{\pm}^*\cdot \eps_{\mp}=1 \ , \quad (\eps_{\pm}^{\mu})^*=\eps_{\mp}^{\mu} \ .
\end{equation}
Let us elaborate on the physical interpretation of the Wigner function components in the circular polarization basis. Since $G^{\mu \nu}$ is invariant under basis rotations, i.e $\eps_{\pm}^{\mu'}\rightarrow e^{\pm i\theta}\eps_{\pm}^{\mu}$, the polarization space components of the Wigner function transform as \cite{Beneke:2010eg}
\begin{equation}
G^{\pm\pm'}\rightarrow G^{\pm\pm} \ , \quad G^{\pm\mp'}\rightarrow e^{\pm 2i\theta}G^{\pm\mp} \ ,
\end{equation}
which reveals that $G^{\pm\pm}$ and $G^{\pm\mp}$ have null $(s=0)$ and integer $(s=\pm 2)$ spin respectively. Moreover, the polarization space components of the Wigner function can be directly related to the Stokes parameters \cite{jackson}, their relation in the circular basis is
\begin{equation}
G^{ab}=
\begin{pmatrix}
G^{++} & G^{+-} \\
G^{-+} & G^{--} 
\end{pmatrix}  
=
\begin{pmatrix}
G^I-G^V & G^Q-iG^U \\
G^Q+iG^U & G^I+G^V 
\end{pmatrix}  
\ .
\end{equation}
Hence, the diagonal components of the Wigner function $G^{\pm\pm}$ relate to the intensity $G^I$ and degree of circular polarization $G^V$ of the photon ensemble, while the off-diagonal components $G^{\pm\mp}$, decomposed into the Stokes parameters $G^Q$ and $G^U$, give information on the polarization phases and are related to the so called $E$ and $B$ polarization modes \cite{Beneke:2010eg}. 

The kinetic equations for each polarization space component of the Wigner function in the circular basis can be simplified after using the identity 
\cite{Nieves:1988qz}
\begin{equation}
\label{circ-basis-identity}
\quad i(\eps_{-}^{*\mu}\eps_-^{\nu}-\eps_+^{*\mu} \eps_{+}^{\nu})=\dfrac{1}{\kappa}\eps^{\mu\nu\alpha\beta}u_\beta q_\alpha  \ ,    
\end{equation}
where we defined $\kappa=\sqrt{(u\cdot q)^2-q^2}$.
Hence, for the diagonal components $G^{++}$ and $G^{--}$, corresponding to right and left handed circularly polarized photons respectively, we find 
\begin{subequations}
\begin{align}
\label{G_++_--_dis}
& \left(q^2 \pm \dfrac{g_{a\g}}{\kappa}\left[(q\cdot\pa a)(u\cdot q)-q^2(u\cdot \pa a) \right] \right)G^{\pm\pm}=0 \ ,
\\
\label{G_++_--_tra}
& iq\cdot \pa \, G^{\pm\pm} = 0 \ .
\end{align}
\end{subequations}
As for the off-diagonal components $G^{+-}$ and $G^{-+}$, reflecting the correlation of different polarization in the photon ensemble, we get
\begin{subequations}
\begin{align}
\label{G_+-_-+_dis}
&  q^2  \, G^{\pm \mp} =0 \ ,
\\
\label{G_+-_-+_tra}
& \left(iq\cdot \pa \pm \dfrac{g_{a\g}}{\kappa}\left[(q\cdot\pa a)(u\cdot q)-q^2(u\cdot \pa a) \right] \right)G^{\pm \mp} =0 \ .
\end{align}  
\end{subequations}
The form of Eqs.~(\ref{G_++_--_dis}-\ref{G_+-_-+_dis}) suggest that the general structure of the Wigner function is
\begin{equation}\label{Wigner_structure}
G^{ab}(X,q)= 4\pi \delta (Q^2_{ab}) \text{sgn}(u\cdot q) f^{ab}(X,q) \ , \quad \lbrace a,b\rbrace =\lbrace +,- \rbrace  \ ,  
\end{equation}
where $\text{sgn}(x)$ is the sign function, the quantities $Q^2_{ab}$ govern the dispersion relations and $f^{ab}(X,q)$ are the off-shell distribution functions for each polarization space component of the Wigner function. The transport equations obeyed by the on-shell distribution functions are derived finding the dispersion relations and imposing the resulting on-shell conditions. Let us show how this is realized in the rest frame of the medium $u^\mu=(1,\boldsymbol{0})$. In this frame, the dispersion relations are obtained as solutions to the equations
\begin{subequations}
\begin{align}\label{dis_R/L}
& Q_{\pm\pm}^2 =\omega^2-\vert\q\vert^2 \pm g_{a\g}\left[\vert\q\vert \pa_0 a+\omega(\hat{\q}\cdot\nabla a) \right]=0 \ , 
\\ \label{dis_EB_modes}
& Q_{\pm\mp}^2=\omega^2-\vert\q\vert^2=0 \ ,
\end{align}    
\end{subequations}
and the transport equations read
\begin{subequations}
\begin{align}
\label{tra_R/L}
& \left(i\omega \pa_0+i \q\cdot \nabla \right)G^{\pm\pm}=0 \ ,
\\
\label{tra_EB_modes}
& \left(i\omega \pa_0+i \q\cdot \nabla\pm g_{a\g}\left[\vert\q\vert \pa_0 a+\omega(\hat{\q}\cdot\nabla a) \right]\right)G^{\pm\mp} = 0 \ ,
\end{align}
\end{subequations}
where we used the notation $q^\mu=(\omega,\q)$ for the photon momentum and defined $\hat{\q}=\q/\vert\q\vert$. Then, we see that the presence of the axion induces a different dispersion relation for the diagonal components of the Wigner function $G^{++}$ and $G^{--}$, while both off-diagonal components $G^{+-}$ and $G^{-+}$ obey a free dispersion relation, unaffected by the axion. Explicitly, solving  Eqs.~(\ref{dis_R/L}-\ref{dis_EB_modes}) we find the following dispersion relations
\begin{subequations}
\begin{align}\label{dis_harari}
& \omega_{\pm\pm}(\q)\approx \vert\q\vert \mp \dfrac{1}{2}g_{a\g} (\pa_0 a + \hat{\q}\cdot\nabla a) \ ,
\\ \label{dis_free}
& \omega_{\pm\mp}(\q)=\vert\q\vert \ .
\end{align}
\end{subequations}
The first equation above coincides with the result first found by Harari and Sikivie \cite{Harari:1992ea}, note that we approximated $\omega_{\pm\pm}(\q)$ at linear order in $g_{a\g}$. There are also negative energy solutions
\begin{subequations}
\begin{align}\label{dis_harari_negative}
& {\widetilde \omega}_{\pm\pm}(\q)\approx -\vert\q\vert \pm \dfrac{1}{2}g_{a\g} (\pa_0 a - \hat{\q}\cdot\nabla a) \ ,
\\ \label{dis_free_negative}
& {\widetilde \omega}_{\pm\mp}(\q)=-\vert\q\vert \ .
\end{align}
\end{subequations}
Imposing the on-shell conditions dictated by Eqs.~(\ref{dis_harari}-\ref{dis_free}) onto Eqs.~(\ref{tra_R/L}-\ref{tra_EB_modes}) respectively leads to the transport equations obeyed by the on-shell distribution functions, that we formally define as
\begin{subequations}
\begin{align} \label{OS_dist_R/L}
& f^{\pm\pm}(X,\q)= f^{\pm\pm}(X,q) \Big\vert_{q_0=\omega_{\pm\pm}(\q)} \ ,
\\   \label{OS_dist_EB_modes}
& f^{\pm\mp}(X,\q)= f^{\pm\mp}(X,q) \Big\vert_{q_0=\vert\q\vert} \ , 
\end{align}
\end{subequations}
for the positive energy solutions. 
Thus, at first order in the gradient expansion and at linear order in $g_{a\g}$ we find
\begin{subequations}
\begin{align} \label{tra_R/L_dist}
& \left(i\pa_0+i\hat{\q}\cdot \nabla \right)f^{\pm\pm}(X,\q) = 0 \ ,
\\ \label{tra_EB_modes_dist}
& \left(i\pa_0+i\hat{\q} \cdot \nabla  \pm g_{a\g}\left(\pa_0 a+\hat{\q}\cdot\nabla a \right) \right)f^{\pm \mp}(X,\q) =0 \ .
\end{align}
\end{subequations}
Please note that the effective velocity appearing in Eq.~\eqref{tra_R/L_dist}, is $\bv_{\text{eff}} = \hat{\q}+\mathcal{O}(\pa_X) $, which consistently ignores the effects of the axion, as it would enter as quantum effect, at second order in the gradient expansion.  Similar distribution functions and transport equations can be defined for the negative energy solutions. 

The conditions we assumed in this work are  equivalent to those carried out in the so called eikonal approximation \cite{Blas:2019qqp}. It has been argued that in the eikonal approximation there is no chiral bending of light in the presence of an axion background in vacuum \cite{Blas:2019qqp}, as the index of refraction of both left and right handed  waves is one in this approximation.  Please note that in this gradient expansion we found that the dispersion law of the right/left handed photons might be also written down as $ ( q \pm \frac{g_{a\g}}{2} \partial a)^2 \approx 0   $, so that the photons travel at the speed of light, as also found in 
\cite{Blas:2019qqp}.  

The  photon current associated to the polarized photons can be defined as
\begin{equation}\label{Current_covariant}
J^{\mu,ab}(X)=(n^{a b}, \, \boldsymbol{j}^{ab}) =\displaystyle\int \dfrac{d^4q}{(2\pi)^4} q^\mu  G^{ab} (X,q) \ .
\end{equation}
In a plasma at thermal equilibrium, and in the frame at rest with the medium, the right and left handed photon distribution function is the Bose-Einstein distribution function $ f_{\rm B} (\omega) = 1/(e^{ \omega /T }-1)$. Then, by direct computation one finds that the difference between equilibrium densities of right and left handed photons is proportional to the temporal variation of the axion field
\begin{equation} \label{diff_densities}
n^{++}-n^{--} =\dfrac{g_{a\g}T^2}{3}\pa_0 a + {\cal O} (g^2_{a\g})\  . 
\end{equation}
Similarly, one finds
\begin{equation}  \label{diff_currents}
\boldsymbol{j}^{++}-\boldsymbol{j}^{--} = \dfrac{g_{a\g}T^2}{9}\nabla a + {\cal O}(g^2_{a\g}) \ , 
\end{equation}
such that the spatial gradient of the axion field induces a difference in the right and left-handed photon currents. On the ohter hand, from Eq.~\eqref{tra_EB_modes_dist} we see that the axion induces a rotation on the $E$ and $B$ modes, if there are such polarizations modes. If $X^\mu_0 = (t_0,\boldsymbol{X}_0) $ denotes an initial coordinate, then one finds that at a final state $X^\mu_f = (t_f,\boldsymbol{X}_f) $
\begin{equation}
f^{\pm \mp}(X_f,\q)= f^{\pm \mp}(X_0,\q) \exp \left\lbrace \mp i g_{a\g}[a (X_f)-a(X_0)]\right\rbrace \ .
\end{equation}
As these components have spin 2, the angle of rotation of these modes is $g_{a\g} (\Delta a)/2$. This accounts for the rotation of the polarization vector first discussed in \cite{Carroll:1989vb,Carroll:1998zi}. In particular, if the initial configuration only contains 
E polarization modes, the axion background induces the appeareance of B-modes. This is an effect that has also already been discussed in the literature \cite{Liu:2006uh,Liu:2016dcg}, and that our transport equation properly encodes.

\section{Collective modes of photons in an axion background}
\label{Sec.III}

The momentum of most quasiparticles that constitute an electromagnetic plasma is of the order of the equilibrium temperature $T$ and/or the chemical potential $\mu$. Collective modes then emerge as perturbations whose typical momentum, that we denote by $Q_{\mu}$, is much lesser than those scales, of the order of the Debye mass $Q_{\mu}\sim m_D\ll T,\mu$. In the previous section, we assumed that the variations of the axion field are much less than the momentum of the photons $\vert\pa_\mu a\vert/\vert a\vert \ll q_\mu \sim T,\mu$, which allowed us to treat the axion as a background field and the photons as quasiparticles. We also emphasized that if the variations of the axion field are of the same order of the photon momentum, then axions should also be treated as quasiparticles. The interaction between the collective modes and the axion field has also to be addressed differently according to the hierarchy between their typical scales. We will assume that the variations of the axion field are much less than the momentum of the collective modes ${\vert\pa_\mu a\vert }/{\vert a\vert} \ll Q_\mu \sim m_D $, so that the axion still can be effectively described as a background field.

In the absence of an axion background or any CP violating effect in the medium, there is a transverse collective mode that is degenerate and a longitudinal collective mode, the so called plasmon, which is absent in vacuum.
The impact of the axion background on the dispersion relations of collective modes is reflected in the dynamics of the dressed propagator $\hat{G}^{\mu\nu}(x,y)$, whose inverse reads in momentum space
\begin{equation}\label{dressed_inv}
\hat{G}^{-1}_{\mu\nu}(Q)= -Q^2 g_{\mu\nu}  +\Pi_{\mu\nu}(Q)+ig_{a\g} \eps_{\mu\nu\alpha\beta}(\pa^\alpha a ) Q^\beta \ .
\end{equation}
Within the frame of reference in which the medium is in motion with velocity $u^{\mu}$, one can establish three independent projectors which are orthogonal to both $Q^\mu$ and $u^\mu$ as \cite{Nieves:1988qz} 
\begin{equation}\label{L/T/P_projectors}
\begin{gathered}
P_T^{\mu\nu}=  \Tilde{g}^{\mu\nu}-P_L^{\mu\nu} \ , \quad P_L^{\mu\nu}=  \dfrac{\Tilde{u}^\mu \Tilde{u}^\nu}{\Tilde{u}^2} \ , \quad P_P^{\mu\nu}= \dfrac{i}{\kappa}\eps^{\mu\nu\alpha\beta}Q_\alpha u_\beta \ ,
\end{gathered}
\end{equation}
where
\begin{equation}\label{quantities}
\Tilde{u}^\mu= \Tilde{g}^{\mu\nu}u_\nu \ , \quad  \Tilde{g}^{\mu\nu}= g^{\mu\nu}-\dfrac{Q^\mu Q^\nu}{Q^2} \ ,   
\end{equation}
and $\kappa$ was given below Eq.~\eqref{circ-basis-identity} (now one should replace $q^\mu\rightarrow Q^\mu$).
The projectors in Eq.~\eqref{L/T/P_projectors} are called transverse, longitudinal and parity odd projectors respectively. They satisfy the properties
\begin{equation}
\begin{gathered}
P_L^2=1 \ , \quad P_T^2= 2 \ , \quad P_P^2=-2  \ ,
\\
P_L^{\mu\nu}P_{T,\mu\nu}=P_L^{\mu\nu}P_{P,\mu\nu}=P_T^{\mu\nu}P_{P,\mu\nu}=0 \ .
\end{gathered}   
\end{equation}
As the effect of a parity odd source, either due to the axion background or the medium, is to split the otherwise degenerate right and left circular polarization modes of the photon \cite{Nieves:1988qz}, it is convenient to introduce additional right $(+)$ and left $(-)$ projectors
\begin{equation}\label{Projectors_P_{+/-}}
P_{h}^{\mu\nu}=\dfrac{1}{2}(P_T^{\mu\nu}+ h P_P^{\mu\nu}) \ , \quad h=\pm  \ ,
\end{equation}
in terms of which, the polarization tensor can be decomposed as 
\begin{equation}\label{Pol_decomposed}
\Pi^{\mu\nu}=\sum_{h=\pm} P^{\mu\nu}_h \left(\Pi_T+h\,\Pi_P\right)-\dfrac{Q^2}{\kappa^2}P^{\mu\nu}_L \Pi_L \ .
\end{equation}
Note that if $\Pi_P=0$ one recovers the usual decomposition of the polarization tensor into its transverse and longitudinal component. The inverse of the dressed propagator in Eq.~\eqref{dressed_inv} can be decomposed in terms of the projectors $P_+^{\mu\nu},P_-^{\mu\nu}$ and $P_L^{\mu\nu}$ too and then inverted, which gives
\begin{equation}
\begin{gathered}\label{Dressed_proj}
\hat{G}^{\mu\nu}(Q)=- \dfrac{\kappa^2}{Q^2}\dfrac{P_{L}^{\mu\nu}}{\kappa^2+\Pi_L}
-\sum_{h=\pm }\dfrac{P_{h}^{\mu\nu}}{Q^2-\Pi_T+h\left(\Pi_P+\dfrac{g_{a\g}}{\kappa}\left[(Q\cdot \pa a)(u\cdot Q)-Q^2(u\cdot \pa a)\right]\right)} 
\ .    
\end{gathered}
\end{equation}
The poles in the dressed propagator above determine the dispersion relations obeyed by the collective modes within the medium. It is worth mentioning that the longitudinal collective mode remains unaffected by the interaction of photons with the axion background. The projectors used to decompose the dressed propagator can be related to the polarization vectors
\begin{equation}\label{Projectors_pol.vectors}
P_+^{\mu\nu}= -\eps_+^{*\mu} \eps_+^{\nu} \ , \quad  P_-^{\mu\nu}= -\eps_-^{*\mu} \eps_-^{\nu} \ , \quad P_L^{\mu\nu}=-\eps_L^\mu \eps_L^\nu \ ,
\end{equation}
where we introduced the longitudinal polarization vector as \cite{Nieves:1988qz}
\begin{equation}\label{Longitudinal_proj}
\eps_L^\mu = \dfrac{\Tilde{u}^\mu}{\sqrt{-\Tilde{u}^2}} \ .
\end{equation}
Thus, we may write the dressed propagator in terms of the polarization basis vectors
\begin{equation}
\begin{gathered}\label{Dressed_pol}
\hat{G}^{\mu\nu}(Q)=\dfrac{\kappa^2}{Q^2}\dfrac{\eps_L^\mu \eps_L^\nu}{\kappa^2+\Pi_L}
+\sum_{h=\pm }\dfrac{\eps_{h}^{*\mu} \eps_{h}^{\nu}}{Q^2-\Pi_T+h\left(\Pi_P+\dfrac{g_{a\g}}{\kappa}\left[(Q\cdot \pa a)(u\cdot Q)-Q^2(u\cdot \pa a)\right]\right)} 
\ .    
\end{gathered}
\end{equation}

Let us now assume a QED plasma at very high temperature, such that one can neglect the electron mass. The Debye mass is then
$m_D^2= e^2T^2/3$ where $e$ is the electron charge so that the momentum of the collective modes is of order $Q_\mu\sim eT$. 
The values of the longitudinal $\Pi_L$ and transverse
$\Pi_T$ polarization tensors components are then well-known, and given by the so called hard thermal loop (HTL) expressions \cite{Bellac:2011kqa}.  Certainly, the axion background also affects the polarization tensor through its coupling to electrons. We ignore those contributions here, but they would be required for a more complete analysis. Then, assuming $\Pi_P=0$ and moving to the rest frame of the plasma $u^\mu=(1,\boldsymbol{0})$, the dispersion relations associated to the two transverse collective modes are obtained as solutions to the equations 
\begin{equation}
\omega^2-\vert\q\vert^2 -\dfrac{m_D^2\omega^2}{2\vert\q\vert^2}\left[1-\left(1-\dfrac{\vert\q\vert^2}{\omega^2}\right)\dfrac{\omega}{2\vert\q\vert}\ln\left(\dfrac{\omega+\vert\q\vert}{\omega-\vert\q\vert}\right)\right]\pm g_{a\g} \left[\vert\q\vert\pa_0 a + \omega(\hat {\q} \cdot\nabla a)\right] =0 \ ,   
\end{equation}
where we used the notation $Q^\mu=(\omega,\q)$.
The effect of the axion background then comes in modifications of order $eT g_{a\g} \pa a$ to the dispersion laws.
While these have to be solved numerically, it is possible to
find simple analytical solutions in some cases. 
For instance, in the long wavelength limit $ m_D \gg \vert\q\vert$ we find the solutions
\begin{equation}\label{frequency_shift}
\omega^2_{\pm}\approx \dfrac{m_D^2}{3}\mp \dfrac{ g_{a\g} m_D}{\sqrt{3}}({\hat{\q}} \cdot \nabla a)\ ,      
\end{equation}
implying that right and left handed circularly polarized collective modes oscillate with different plasma frequencies, and at this expansion order also different 
from that of the plasmon mode, which is $\omega_L=m_D/\sqrt{3}\equiv \omega_{\text{pl}}$ \cite{Bellac:2011kqa}. On the other hand, in the regime $m_D\ll\vert\q\vert\ll T$ one finds
\begin{equation}\label{mass_shift}
\omega^2_{\pm}\approx \vert\q\vert^2+m_D^2\mp  g_{a\g}(\pa_0 a+{\hat{\q}} \cdot \nabla a)  \ ,  
\end{equation}
so that the axion produces a different shift on the effective asymptotic  masses for right and left handed modes. It is also interesting to study the limit $\omega \ll \vert\q\vert$, as in this regime, and due to Landau damping, 
there is an additional family of poles in the transverse modes which are purely imaginary. Let us assume without loss of generality that $\Im(\omega)>0$ where $\Im$ denotes the imaginary part, then one finds the solutions
\begin{equation}
\omega_{\pm} = -i \frac{4 \vert \q \vert ^3}{\pi m^2_D} \left( 1 \mp \frac{ g_{a\g} \pa_0 a}{\vert \q \vert} \right)=-i\gamma_{\pm}\ ,
\end{equation}
This has the same form of the chiral instabilities that  are found in chiral media characterized by an imbalance in the population of right and left handed fermions \cite{Akamatsu:2013pjd}. In fact, it has the same form, after identifying the chiral chemical potential $\mu_5$ with $ g_{a\g} \pa_0 a/2\alpha$ \cite{Akamatsu:2013pjd}, \cite{Carignano:2018thu}.  The collective modes evolve in time
as $\exp\left(- i\omega_{\pm}t\right)\sim \exp\left(-\gamma_{\pm}t\right)$, and they would become unstable if $\gamma_{\pm}$ becomes negative. As the sign of $\pa_0 a$ can be either positive or negative, this leads to the conditions $\pm g_{a\g}\pa_0 a > \vert\q\vert$, however, in this article we assumed that ${\vert\pa_\mu a\vert }/{\vert a\vert} \ll Q_\mu $, as the axion is treated as a background field. Therefore, taking this assumption into account and due to the smallness of the axion-photon coupling constant $g_{a\g}$, we conclude that $\gamma_{\pm}$ remains positive and there are no unstable modes in this case. 

\section{ Discussion} 
\label{Sec.IV}	

We have developed a quantum kinetic theory in the collisionless limit for photons with the presence of an axion background, which is summarized in Eqs.(\ref{Kinetic_Equations}-\ref{gauge_ambiguity}). Performing a gradient expansion of the operators and the Wigner function (or phase-space distribution), we derived their classical limit and projected the resulting equations on a basis of polarization vectors, yielding Eqs.(\ref{classical_KE}-\ref{classical-Lorentzcondition}). A considerable advantage of this last projection is that the components of the Wigner function in polarization space can be directly related to the Stokes parameters, thus having a clear physical interpretation. Then, using a circular polarization basis of vectors, we derived the transport equations obeyed by the on-shell distribution functions $f^{\pm\pm}$ and $f^{\pm\mp}$ in the rest frame of the medium, given by Eqs.(\ref{tra_R/L_dist}-\ref{tra_EB_modes_dist}) respectively, which is the central result of this article. Those equations properly encode features of axion electrodynamics, such that right and left handed circular polarized photons obey different dispersion relations, or the phase rotation of the polarization plane, which have been explored before in the literature.

A similar transport equation for photons in an axion background was derived in Ref.\cite{Shakeri:2022usk}, there the authors derived a transport equation for the Stokes parameters in a time-dependent axion background. Our treatment is more general and fully covariant, and allows for the incorporation of different sort of corrections. For the comparison with Ref.\cite{Shakeri:2022usk} see Appendix \hyperref[B]{B}.

There are several ways in which our work could be extended, for instance, we could include the effects of collisions of photons with the quasiparticles of the thermal bath or compute quantum corrections to our classical transport equations. Another interesting generalization would be to consider that photons propagate through a non flat space time, as considered in 
 Ref.\cite{Beneke:2010eg}, as the presence of the axion background could provide new sources of $B$ mode polarization. 

We have also addressed the effects of the axion background on the photon collective oscillations within the medium through the photon-axion interaction. As expected, the axion background breaks the degeneracy of right and left handed circular polarized collective modes, while the plasmon remains unaffected. The contribution of the axion background is of order $eg_{a\g}T(\pa a)$, where $\pa a$ is the scale associated to the variations of the axion field. We also considered limiting cases for the dispersion relations of the transverse collective modes; in the regime $m_D \gg \vert\q\vert$ the axion produces a shift on the oscillation frequencies of right and left handed polarized collective modes, see Eq.\eqref{frequency_shift}, while in the regime $m_D \ll \vert\q\vert\ll T$ the axion modifies their effective asymptotic masses (cf. Eq.\eqref{mass_shift}). 

It has been argued that specific photon modes in chiral media may not propagate and experience instabilities \cite{Qiu:2016hzd}. However, under the considerations of this work, we have shown that if the chiral media consists of an axion background, both right and left handed photons are propagating modes, since the assumption that the axion field acts as a background prevents those instabilities to occur. We stress that the situation would change if the axions are considered as quanta, interacting with photons through the Lagrangian of Eq.\eqref{L_photon+axion}. A similar reasoning and conclusion applies for the collective modes, that we have elaborated in Sec.\hyperref[Sec.II]{II}.

A relevant scenario, that we have not considered in this work, is when the variations of the axion field are comparable to the momentum scale of the collective modes, as in that case interactions between axions and collective modes can occur, leading to interesting phenomena and possible windows for detection \cite{Raffelt:1987np,Mikheev:2009zz,Caputo:2020quz,Redell:2016alb}.


\section*{Acknowledgements} 
\label{ack}		

We thank Diego Blas for discussions. This work was supported by Ministerio de Ciencia, Investigaci\'on y Universidades (Spain) MCIN/AEI/10.13039/501100011033/ FEDER, UE, under the projects PID2019-110165GB-I00 and PID2022-139427NB-I00, by Generalitat de Catalunya by the project 2021-SGR-171 (Catalonia). This work was also partly supported by the Spanish program Unidad de Excelencia
 Maria de Maeztu CEX2020-001058-M, financed by MCIN/AEI/10.13039/501100011033.

\appendix
\section{Kinetic equations}
\label{A}
In this section we give some details on how to derive the kinetic equations Eqs.~(\ref{Dispersion}-\ref{Transport}). We start by defining the Wigner transform of the sum and difference of the equations of motion given by Eqs.~(\ref{motion_x}-\ref{motion_y}) as
\begin{equation}
\begin{gathered}
(I_{\pm})_{\lambda\nu}=\int d^4 s\, e^{i q \cdot s} \big\lbrace \left(g_{\mu\lambda}\square-g_{a\g}\eps_{\mu\lambda \alpha\beta}(\pa^\alpha a)\pa^\beta\right)_x G^{\mu}_{\  \nu}(x,y) \\
 \pm \left(g_{\mu\nu}\square-g_{a\g}\eps_{\mu \nu \alpha\beta}(\pa^\alpha a)\pa^\beta\right)_y  G^{\  \mu}_{ \lambda}(x,y) \big\rbrace \ . 
\end{gathered}
\end{equation}
Then, we move from configuration space variables $x^\mu$ and $y^\mu$ to Wigner space variables $X^\mu$ and $s^\mu$ using the relations
\begin{subequations}
\begin{gather}
X^\mu=\dfrac{x^\mu+y^\mu}{2} \ , \quad s^\mu=x^\mu-y^\mu \ ,
\\
\pa_x^\mu =\dfrac{1}{2}\pa_X^\mu+\pa_s^\mu \ , \quad
\pa_y^\mu =\dfrac{1}{2}\pa_X^\mu-\pa_s^\mu \ .
\end{gather}
\end{subequations}
Doing so, we can write
\begin{equation}
\begin{gathered}
(I_{\pm})_{\lambda\nu}=\int d^4 s\, e^{i q \cdot s} 
\bigg\lbrace \bigg[g_{\mu\lambda}\left(\pa_s\cdot\pa_X+\pa_s^2+\dfrac{1}{4}\pa_X^2\right)
-g_{a\g}\eps_{\mu\lambda \alpha\beta}A^{\alpha\beta}(X,s)\bigg] \, G^{\mu}_{\ \nu}(X+s/2,X-s/2) 
\\
\pm \bigg[g_{\mu\nu}\left(-\pa_s\cdot\pa_X+\pa_s^2+\dfrac{1}{4}\pa_X^2\right)
-g_{a\g}\eps_{\mu\nu \alpha\beta}A^{\alpha\beta}(X,-s)\bigg] \, G^{\ \mu}_{\lambda}(X+s/2,X-s/2) \bigg\rbrace \ . 
\end{gathered}
\end{equation}
Where we defined the following operator, acting on the Wigner function
\begin{equation}
A^{\alpha\beta}(X,s)=\left(\dfrac{1}{2}\pa^\alpha_X +\pa_s^\alpha\right) a(X+s/2)\left(\dfrac{1}{2}\pa^\beta_X +\pa_s^\beta\right) \ .   
\end{equation}
Now we perform a gradient expansion of the axion field 
\begin{equation}
a(X+s/2)= \sum_{n=0}^{\infty} \dfrac{1}{n!} \left(\dfrac{s\cdot\pa_X}{2}\right)^n a(X)\ ,
\end{equation}
and after the Wigner transformation, we find
\begin{equation}
\begin{gathered}
(I_{\pm})_{\lambda\nu}=
\bigg\lbrace \bigg[g_{\mu\lambda}\left(-iq\cdot \pa_X-q^2+\dfrac{1}{4}\pa_X^2\right)
-g_{a\g}\eps_{\mu\lambda \alpha\beta}A^{\alpha\beta}(X,q)\bigg] \, G^{\mu}_{\ \nu}(X,q) 
\\
\pm \bigg[g_{\mu\nu}\left(iq\cdot \pa_X-q^2+\dfrac{1}{4}\pa_X^2\right)
-g_{a\g}\eps_{\mu\nu \alpha\beta}A^{*,\alpha\beta}(X,q)\bigg] \, G^{\ \mu}_{\lambda}(X,q) \bigg\rbrace \ . 
\end{gathered}
\end{equation}
Where now $A^{\alpha\beta}(X,q)$ is given by Eq.\eqref{A_tensor}, in which we neglected the arguments for simplicity. So finally, we find for the dispersion relation
\begin{equation}\label{I_+}
\begin{gathered}
(I_{+})_{\lambda\nu}=
\left(-2q^2+\dfrac{1}{2}\pa_X^2\right) G_{\lambda\nu}(X,q)
\\
-g_{a\g}\eps_{\mu\lambda \alpha\beta}A^{\alpha\beta}(X,q)G^{\mu}_{\ \nu}(X,q) 
-g_{a\g}\eps_{\mu\nu\alpha\beta}A^{*,\alpha\beta}(X,q) \, G^{\ \mu}_{ \lambda}(X,q) =0 \ .
\end{gathered}
\end{equation}
While for the transport equation
\begin{equation}\label{I_-}
\begin{gathered}
(I_{-})_{\lambda\nu}=
 -2i(q\cdot\pa_X) G_{\lambda\nu}(X,q)
\\
-g_{a\g}\eps_{\mu\lambda \alpha\beta}A^{\alpha\beta}(X,q)G^{\mu}_{\ \nu}(X,q) 
+g_{a\g}\eps_{\mu\nu \alpha\beta}A^{*,\alpha\beta}(X,q) \, G^{\ \mu}_{ \lambda}(X,q) =0 \ .
\end{gathered}
\end{equation}
Now we divide both Eqs.~(\ref{I_+}-\ref{I_-}) by a factor of $-2$, also rising the indices in the Wigner function and the Levi-Civita tensor using the property $\eps_{\mu\nu\rho\sigma}=-\eps^{\mu\nu\rho\sigma}$ we reach to
\begin{equation}
\begin{gathered}
\left(q^2-\dfrac{1}{4}\pa_X^2\right) G^{\lambda\nu}
-\dfrac{g_{a\g}}{2}\left(\eps^{\mu\lambda \alpha\beta}A_{\alpha\beta}G_{\mu}^{\ \nu}
+\eps^{\mu\nu\alpha\beta}A^*_{\alpha\beta} \, G^{\lambda }_{\ \mu} \right)=0 \ ,
\\
\left(iq\cdot \pa_X\right) G^{\lambda\nu}
-\dfrac{g_{a\g}}{2}\left(\eps^{\mu\lambda\alpha\beta}A_{\alpha\beta}G_{\mu}^{\ \nu}
-\eps^{\mu\nu \alpha\beta}A^*_{\alpha\beta} \, G^{\lambda }_{\ \mu} \right)=0 \ ,
\end{gathered}
\end{equation}
which exactly give Eqs.~(\ref{Dispersion}-\ref{Transport}) after relabeling some indices and using the antisymmetry property of the Levi-Civita tensor.

\section{Classical transport equation in the linear polarization basis}
\label{B}
In this section we give the classical kinetic Eqs.~(\ref{Dispersion-classical}-\ref{Transport-classical}) in a linear polarization basis and then write them in terms of the Stokes parameters. We start by introducing a linear polarization basis vectors $\eps_a^{\mu}=\lbrace \eps_{1}^{\mu},\eps_{2}^\mu\rbrace$, satisfying $\eps_{a}^*\cdot \eps_{b}=\delta_{ab}$ and $ (\eps_{a}^{\mu})^*=\eps_{a}^{\mu}$.
The relation between the components of the Wigner function and the Stokes parameters in a linear polarization basis is \cite{jackson}
\begin{equation}
G^{ab}=
\begin{pmatrix}
G^{11} & G^{12} \\
G^{21} & G^{22} 
\end{pmatrix}  
=
\begin{pmatrix}
G^I+G^Q & G^U-iG^V \\
G^U+iG^V & G^I-G^Q 
\end{pmatrix}  
\ ,
\end{equation}
so that, after using the identity
\begin{equation}
\label{linear-basis-identity}
\quad \eps_{1}^{\mu}\eps_2^{\nu}-\eps_2^{\mu} \eps_{1}^{\nu}=\dfrac{1}{\kappa}\eps^{\mu\nu\alpha\beta}u_\beta q_\alpha  \ ,    
\end{equation}
the dispersion laws for the Stokes parameters may be written as
\begin{subequations}
\begin{align}
& q^2 G^I -\dfrac{g_{a\g}}{\kappa}\left[(q\cdot\pa a)(u\cdot q)-q^2(u\cdot \pa a) \right] G^V=0 \ , \\
& q^2 G^V-\dfrac{g_{a\g}}{\kappa}\left[(q\cdot\pa a)(u\cdot q)-q^2(u\cdot \pa a) \right] G^I=0 \ , \\
& q^2 G^Q =0 \ , \\
& q^2 G^U =0 \ . 
\end{align}   
\end{subequations}
A problem with this formulation, that was not discussed in Ref.~\cite{Shakeri:2022usk} is that the equations for $G^I$ and $G^V$ are coupled, so that they can not be treated individually as propagating modes. Instead, the propagating modes are the combinations $G^I\mp G^V$, corresponding to right and left handed photons respectively, whose dispersion relations can be obtained by solving
\begin{equation}
\left( q^2 \pm \dfrac{g_{a\g}}{\kappa}\left[(q\cdot\pa a)(u\cdot q)-q^2(u\cdot \pa a) \right] \right)\left(G^I\mp G^V\right)=0 \ ,
\end{equation}
which yields to the solutions of Eqs.~(\ref{dis_harari}-\ref{dis_harari_negative}) in the rest frame. The transport equations in the linear polarization basis read
\begin{subequations}
\begin{align}
& (q\cdot \pa) \left(G^I\mp G^V\right) =0 \ , \\
& (q\cdot \pa) G^Q+ \dfrac{g_{a\g}}{\kappa}\left[(q\cdot\pa a)(u\cdot q)-q^2(u\cdot \pa a) \right] G^U=0 \ , \\
& (q\cdot \pa) G^U - \dfrac{g_{a\g}}{\kappa}\left[(q\cdot\pa a)(u\cdot q)-q^2(u\cdot \pa a) \right] G^Q=0 \ .
\end{align}   
\end{subequations}
The coupled equations for $G^Q$ and $G^U$ can be solved by applying the operator $(q\cdot\pa)$ on each equation and neglecting terms with two derivatives acting on the axion field. Thus, we find
\begin{subequations}
\begin{align}
&  (q\cdot \pa)^2G^Q+ \dfrac{g_{a\g}^2}{\kappa^2}\left[(q\cdot\pa a)(u\cdot q)-q^2(u\cdot \pa a) \right]^2 G^Q =0 \ , \\
& (q\cdot \pa)^2 G^U-\dfrac{g_{a\g}^2}{\kappa^2}\left[(q\cdot\pa a)(u\cdot q)-q^2(u\cdot \pa a) \right]^2 G^U =0 \ .
\end{align}
\end{subequations}
Now, the general structure of the Stokes parameters is $G^{Q,U}(X,q)=4\pi  \delta(q^2)\text{sgn}(u\cdot q)f^{Q,U}(X,q)$, being $f^{Q,U}(X,q)$ the corresponding off-shell distribution functions. Moving to the rest frame and imposing the on-shell condition, we reach to
\begin{subequations}
	\begin{align}
		&  \left[(v\cdot \pa)^2+ \Omega^2\right] f^Q (X,\q) =0 \ , \\
		& \left[(v\cdot \pa)^2 -\Omega^2\right]  f^U(X,\q) =0 \ ,
	\end{align}
\end{subequations}
where $f^{Q,U}(X,\q)$ are the on-shell distribution functions for the positive energy solutions, also we introduced the velocity vector $v^\mu=(1,\hat{\q})$ and the frequency $\Omega=g_{a\g}(v\cdot \pa a)$. Note that $\Omega$ coincides with the frequency defined in Ref.~\cite{Shakeri:2022usk} when neglecting the gradient of the axion field (there is a difference of a factor of $2$ due to our distinct definition of the dual of the electromagnetic tensor $\Tilde{F}^{\mu\nu}$).

\end{document}